\documentstyle[aps,preprint]{revtex} \baselineskip 12pt
\begin{document}
\title{ Quantum gauge symmetry from finite field dependent BRST 
transformations} 

\vspace{.4in}

\author{Rabin Banerjee \footnote{ Email: rabin@boson.bose.res.in}
 \ and\ \ \ Bhabani Prasad Mandal \footnote {Email: bpm@boson.bose.res.in}
}

\address{
 S.N. Bose National Center for Basic Sciences,\\
Block-JD; Sector-III; Salt Lake ,\\
 Calcutta-700 091, India.\\
}

\vspace{.4in}

\maketitle

\vspace{.4in}

\begin{abstract}

Using the technique of finite field dependent BRST transformations we
show that the classical massive Yang-Mills  theory and the pure Yang-Mills
 theory whose gauge symmetry is broken by a gauge fixing term are 
identical from the view point of quantum gauge symmetry. The explicit 
infinitesimal transformations which leave the massive Yang-Mills theory 
BRST invariant are given.
\end{abstract}

\newpage
 
In a recent paper \cite{fuj} it was shown that a classical massive gauge
theory does not have an essential difference, at the quantum level, from a
gauge invariant theory whose gauge symmetry is broken by a gauge fixing 
term. Specifically, the classical lagrangians,
\begin{equation} 
{\cal L} = {\cal L} _{YM} - \frac{ m^2}{2} A_\mu^a A_\mu^a
\label{em}  
\end{equation}
and
\begin{equation}
{\cal L} = {\cal L} _{YM} - \frac{1}{2} (\partial _\mu A_\mu^a)^2 
\label{egf} 
\end{equation}
where  $ {\cal L} _{YM} $ is the Yang-Mills lagrangian,
\begin{equation}
{\cal L} _{YM} = - \frac{ 1}{4} (\partial _\mu A_\nu ^a -\partial _\nu A_\mu
^a +gf^{abc} A_\mu ^b A_\nu ^c)^2
\label{eym}
\end{equation}
could be given an identical physical meaning, both representing an effective
gauge fixed lagrangian associated with the quantum theory defined by
\begin{equation}
\int {\cal D} A_\mu ^a {\cal D} B^a {\cal D} \bar{c} ^a {\cal D} c^a 
\exp \left \{ -S_{YM} +\int d^4 x \left [-iB^a(\partial ^\mu A_\mu^a)
  + \bar{c}^a (-\partial _\mu(D^\mu c)^a) \right ] \right \}
\label{44}
\end{equation}
that is invariant under the  BRST transformations,
\begin{eqnarray}
\delta A_\mu ^a &=& -i(D_\mu c )^a\ \  \delta \lambda \nonumber \\  
\delta c^a &=& - i\frac{ g}{2} f^{abc} c^b c^c \ \ \delta \lambda \nonumber
\\ 
\delta \bar{c} ^a &=& B ^a \ \ \delta \lambda \nonumber \\  
\delta B^a &=& 0 
\label{ibrs}
\end{eqnarray}
where $ \delta \lambda $ is an infinitesimal Grassmann parameter.  
This is to be contrasted with the conventional interpretation of regarding
(\ref{em}) as a massive vector theory and (\ref{egf}) as an effective
Yang-Mills theory with a covariant gauge fixing term.

In this paper we shall 
show the equivalence of the quantum theories defined by (\ref{em}) and
(\ref{egf}) by following the method of finite field dependent BRST (FFBRST)
 transformations developed by
one of us \cite{jm}. In particular this method, which will be briefly
reviewed below, connects quantum gauge theories in different gauges.
Here we start from the conventional gauge fixed
Yang-Mills lagrangian defined by (\ref{egf}). The explicit FFBRST  
transformations are then stated which maps this theory to one whose
lagrangian is defined by (\ref{em}), thereby showing the connection between
them. We also get the form of the transformations that preserve the BRST
invariance of the quantum theory defined by (\ref{em}). Finally we suggest
a possible connection between our approach and that adopted in \cite{fuj},
which was based on a modified quantization scheme \cite{zpj,npb}, where the
variation of the gauge field in the path integral is taken over the entire
gauge orbit.

Let us now briefly review the FFBRST approach \cite{jm,sdj,sdj0,sdj1}.
 FFBRST transformations are
obtained by an integration of infinitesimal  (field dependent ) BRST
transformations  \cite{jm}. In this method all the fields are function
of some parameter, $ \kappa : 0\le \kappa \le 1$. For a generic field $ \phi (x, \kappa),\
\phi(x, \kappa =0 ) = \phi(x) $ and $ \phi(x, \kappa=1) = \phi ^\prime
(x).$ Then the infinitesimal field dependent BRST transformations are 
defined as,
\begin{equation}
\frac{ d}{d \kappa}\phi(x, \kappa ) = \delta _{BRST } \phi(x, \kappa )
\Theta ^ \prime [\phi(x,\kappa )]
\label{ibr}
\end{equation}
where $\Theta ^\prime d \kappa $ is an infinitesimal field dependent parameter.
It has been shown by integrating these equations from $ \kappa=0$ to $\kappa=1$ 
that $\phi^\prime ( x) $ are related to $\phi(x) 
$ by FFBRST,
\begin{equation}
\phi^\prime (x) = \phi(x) + \delta _{BRST}\phi(x) \Theta [\phi(x)]
\label{fbrs}
\end{equation}
where $ \Theta [\phi(x)] $ is obtained from $\Theta ^\prime [\phi(x)] $   
through the relation,
\begin{equation}
\Theta [\phi(x)] = \Theta ^\prime [\phi(x)] \frac{ \exp f[\phi(x)]
-1}{f[\phi(x)]}
\label{80}
\end{equation}
and $f$ is given by $ f= \sum_i \frac{ \delta \Theta ^\prime (x)}{\delta
\phi_i(x)} \delta _{BRST}\phi_i(x) $

The choice of the parameter $ \Theta ^\prime $ is crucial in connecting different
effective gauge theories by means of the FFBRST.
In particular the FFBRST of Eq. (\ref{fbrs}) with
$ \Theta ^\prime [\phi(x,\kappa)] = i \int \bar{c} ^a (y) \left [
F^a [A( \kappa )] - F^{\prime a }[ A(\kappa)] \right ] $
relates the Yang-Mills theory with an arbitrary gauge fixing $F[A]$ to the
Yang-Mills theory with another arbitrary gauge fixing $F^\prime [A]$ \cite{sdj}.  

The meaning of these field transformations is as follows. We consider
the vacuum expectation value of a gauge invariant functional $G[\phi]$
in some arbitrary gauge $F[A]$,
\begin{equation}
<< G[\phi]>> \equiv \int {\cal D} \phi G[\phi] \exp(iS^F_{eff}[\phi])
\label{90}
\end{equation}
where,
\begin{equation}
S_{eff}^F = S_0 - \frac{ 1}{2 }\int d^4x F^2[A] -\int d^4 x \bar{c} ^a
W^{ab} c^b
\label{sf} 
\end{equation}
with
\begin{equation}
W^{ab} = \frac{ \delta F^a}{\delta A_\mu ^c} D^{cb}_\mu [A]
\label{fpd}
\end{equation}
Here $S_0$  is the pure Yang-Mills action obtained from (\ref{eym})
 and the covariant derivative, $ D^{ab}_\mu [A] \equiv \delta ^{ab}\partial
_\mu + g f^{abc}A^c_\mu $. For simplicity we have set the gauge parameter
$\lambda =1 $ in the gauge fixing term $ \frac{ 1}{2 \lambda }\int d^4x
F^2[A] $.

Now we perform the FFBRST transformations $\phi\rightarrow \phi^\prime 
 $ given by (\ref{fbrs}). We have then
\begin{equation}
<<G[\phi]>>=  <<G[\phi^\prime ]>> = \int {\cal D} \phi ^\prime
J[\phi^\prime ] G[\phi^\prime ] \exp(iS^F_{eff}[\phi^\prime ])
\label{def}
\end{equation}
on account of BRST invariance of $S_{eff}^F$ and gauge invariance of $G[\phi]$.
Here $J[\phi^\prime ]$ is the Jacobian associated with FFBRST and defined
as,
\begin{equation}
{\cal D} \phi = {\cal D} \phi^\prime J[\phi^\prime ]
\label{jac} 
\end{equation}

As shown in \cite{jm} for the special case $G[\phi]=1$ the Jacobian
$ J[\phi^\prime ]$ in Eq (\ref{def}) can always be replaced by
$ \exp(iS_1[\phi^\prime ]$) with,
\begin{equation}
S_{eff}^F[\phi^\prime ] +S_1[\phi^\prime ] = S_{eff}^{F^\prime
}[\phi^\prime]
\label{s1}
\end{equation}
where
\begin{equation}
S_{eff}^{F^\prime } = S_0 - \frac{ 1}{2 }\int d^4x F^{\prime 2 }[A] -
\int d^4 x \bar{c} ^a W^{\prime ab} c^b
\label{sf1}
\end{equation}
with
\begin{equation}
W^{\prime ab} = \frac{ \delta F^{\prime a}}{\delta A_\mu ^c} D^{cb}_\mu [A]
\end{equation}

The extra piece in the action which
arises from the Jacobian of such FFBRST is given by,
\begin{equation}
S_1[\phi] = \int d^4 x \left [ - \frac{ 1}{2  } F^{\prime 2}[A] +
\frac{1}{2 } F^2[A] + \bar{c} [W-W^\prime ] c \right ]
\label{s2}
\end{equation}
Thus the FFBRST in Eq. (\ref{fbrs}) takes the theory with gauge $F$ to 
the corresponding theory with gauge $F^\prime $.

We are now ready to apply this machinery to the present problem.
We start with the generating functional for the Yang-Mills theory in
the Lorentz gauge,
\begin{equation}
Z=\int {\cal D} A_\mu {\cal D} c {\cal D} \bar{c} \exp(iS_{eff}^L )
\label{zl}
\end{equation}
where
\begin{equation}
S^L_{eff}= S_0 - \frac{ 1}{2}\int d^4x (\partial ^\mu A_\mu )^2-\int d^4x
\bar{c} W c
\label{tt}
\end{equation}
with $ W= \partial ^\mu D_\mu $ is the Faddeev-Popov determinant.
We now apply FFBRST [Eq. (\ref{fbrs}) ] with 
\begin{equation}
\Theta ^\prime = i\int d^4 y \bar{c} ^a (y)\left [ \partial ^\mu A_\mu^a - m 
\frac{ \omega ^\mu}{|\omega |} A_\mu^a \right ](y)
\label{tt1}
\end{equation}
where $\omega ^\mu  $ is an arbitrary 4-vector,
to the expression for the generating functional to obtain,
\begin{equation}
Z=\int {\cal D} A_\mu^\prime  {\cal D} c^\prime  {\cal D} \bar{c}^\prime 
 \exp i(S_{eff}^L + S_1)
\label{tt2}
\end{equation}
The additional piece in the action comes from the non-trivial Jacobian
of the FFBRST and can be written using Eq (\ref{s2})
\begin{equation}
S_1= \int d^4 x \left [ - \frac{ 1}{2|\omega| ^2} m^2 (\omega ^\mu A_\mu)^2 + 
\frac{ 1}{2} (\partial ^\mu A_\mu)^2 -\bar{c} (W^\prime -W)c \right ]
\label{tt3}
\end{equation}
with $W^\prime = m\frac{  \omega ^\mu}{|\omega |} D_\mu $.
Hence we obtain the generating functional for a new effective action given by,
\begin{equation}
S_{eff} = S_0 -\int d^4x \left [ \frac{ 1}{2} A_\mu M^{\mu\nu}A_\nu +
\bar{c} m \frac{ \omega ^\mu}{|\omega |} D_\mu c \right ]
\label{final}
\end{equation}

where $M^{\mu\nu}$ is a generalized mass matrix,
\begin{equation}
M^{\mu\nu} = m^2\frac{ \omega^\mu \omega ^\nu}{|\omega |^2}
\label{tt4}
\end{equation}
This effective action (\ref{final})  corresponds to the 
 Yang-Mills lagrangian with a generalized
mass term.
It shows the connection between the Lorentz gauge and a generalized
 `mass' gauge $ \frac{ 1}{2} A_\mu M^{\mu\nu} A_\nu $ in the context of
Yang-Mills
theory. To exactly reproduce the familiar mass term, we restrict the
arbitrary vector $\omega ^\mu$  to be of infinitesimal form satisfying the
symmetric  multiplication rule,
\begin{equation}
\frac{ \omega^\mu \omega _\nu}{|\omega| ^2} = \frac{ g^\mu_\nu}{4}
\label{tt5}
\end{equation}
In that case the gauge fixing term is $\frac{ 1}{8} m^2 A_\mu A^\nu$ 
which coincides with the standard mass term, after a proper normalization
of $m$.

The infinitesimal BRST transformations which leave the Yang-Mills 
 theory with a mass term (\ref{final}) invariant 
are given by
\begin{eqnarray}
\delta A_\mu^a &=& D_\mu^{ab} c^b\ \  \delta \lambda \nonumber \\ 
\delta c^a &=& - \frac{ g}{2}f^{abc}c^bc^c\ \  \delta \lambda \nonumber \\ 
\delta \bar{c}^a &=& m \frac{ \omega ^\mu}{|\omega |} A_\mu^a \ \ 
\delta \lambda 
\end{eqnarray}

We have shown how, by means of finite field dependent BRST transformations,
it was possible to interpolate between the Yang-Mills theory in the
covariant gauge to the Yang-Mills theory in a mass like gauge. Since FFBRST
also connects the Yang-Mills theory in the axial and covariant gauges
\cite{sdj,sdj0} it is clear that the Yang-Mills theory with a mass like gauge
fixing term can also be obtained from other starting points. In this paper
we took the covariant gauge as the starting point for reasons  of convenience
and also comparing our analysis with \cite{fuj}. It may be pointed out that
the latter approach is based on the variation of the gauge variable along
the entire gauge orbit, without taking any specific limit of the gauge
fixing parameter. Consequently there seems to be a connection
between this approach and the FFBRST method, which is not
altogether surprising. Carrying out the integration over the complete gauge
orbit would be simulated by  finite BRST transformations instead of the 
conventional infinitesimal one. We feel it might be useful to pursue this
connection in a later work. 

\end{document}